\documentclass[twocolumn, pra, showpacs,superscriptaddress]{revtex4}
\usepackage{amssymb}
\usepackage{graphicx}
\usepackage{dcolumn}
\usepackage{bm}
\usepackage{amsmath}

\begin{document}

\title{Dynamical evolutions in non-Hermitian triple-well system with complex
potential}
\author{Liping Guo}
\email{guolp@sxu.edu.cn}
\affiliation{Institute of Theoretical Physics, Shanxi University, Taiyuan 030006, P. R.
China}
\author{Lei Du}
\affiliation{Institute of Theoretical Physics, Shanxi University, Taiyuan 030006, P. R.
China}
\author{Chuanhao Yin}
\affiliation{Institute of Physics, Chinese Academy of Sciences, Beijing 100080, P. R.
China}
\author{Yunbo Zhang}
\affiliation{Institute of Theoretical Physics, Shanxi University, Taiyuan 030006, P. R.
China}
\author{Shu Chen}
\affiliation{Institute of Physics, Chinese Academy of Sciences, Beijing 100080, P. R.
China}

\begin{abstract}
We investigate the dynamical properties for non-Hermitian triple-well system
with a loss in the middle well. When chemical potentials in two end wells
are uniform and nonlinear interactions are neglected, there always exists a
dark state, whose eigenenergy becomes zero, and the projections onto which
do not change over time and the loss factor. The increasing of loss factor
only makes the damping form from the oscillating decay to over-damping
decay. However, when the nonlinear interaction is introduced, even
interactions in the two end wells are also uniform, the projection of the
dark state will be obviously diminished. Simultaneously the increasing of
loss factor will also aggravate the loss. In this process the interaction in
the middle well plays no role. When two chemical potentials or interactions
in two end wells are not uniform all disappear with time. In addition, when
we extend the triple-well system to a general $(2n+1)$-well, the loss is
reduced greatly by the factor $1/2n$ in the absence of the nonlinear
interaction.
\end{abstract}

\startpage{1}
\endpage{6}
\maketitle

\section{Introduction}

In quantum mechanical pictures Hamiltonian must be Hermitian to describe a
physical system, which is sufficient to ensure that the system has real
energy eigenvalues and the conservation of the number of particles. But this
condition is too rigorous in real systems. In optics, a non-Hermitian
Hamiltonian is used to describe the propagation of light in the medium with
complex refraction index \cite%
{Moiseyev,Longhi,Kivshar,Luo1,Musslimani,Kottos}. Recently the controlled
removal of atoms from a Bose-Einstein condensate (BEC) can be realized by a
narrow electron beam or a narrow laser beam \cite{Ott,Konotop}, which
promotes simulations of the atomic system with dissipation. The systems with
dissipation process are described via the non-Hermitian Hamiltonians with
negative imaginary chemical potential \cite%
{Bender,Bender2,Mostafazadeh,Chen,Wunner,Niederle,Livi,Ng,Konotop1} and can
be solved in terms of the master equations \cite%
{Zoller,Zoller2,Cirac,Wimberger,Altman,Ciuti,Lee}. More importantly, in most
cases dissipation is considered as an undesirable destructing factor, thus
people make arduous efforts to avoid it at all if possible, such as
inverting the dissipation by means of an intrinsic mechanism to balance the
losses \cite{Konotop,Rotter,Wunner1}, probing a quantum system with
controlled dissipation \cite{Ivana}, and designing the effective dissipative
process in an optical superlattice using the coupling between the system and
the reservoir \cite{Zoller}, etc. Massive efforts have been invested in the
study of the dynamics of non-Hermitian systems in experiment and theory \cite%
{Zoller,Greiner,Bloch,Lee2,Cao}.

The study of few-well systems reveals a variety of interesting quantum
phenomena. For example, condensates in double or three wells have popularly
been investigated both theoretically and experimentally \cite%
{Gati,Smerzi1,LiuJie1,LiuJie2,Oberthaler1,Oberthaler2,Oberthaler3,Levy,Smerzi2,Smerzi,Shore,zhu,fu,Munro}%
. In past years nonlinear Josephson oscillation and self-trapping phenomena
are two of many important findings for double wells. However, more
attentions have been focused on three-well system \cite{Shore,zhu,fu,Munro},
which has more abundant physical picture by adjusting the tunneling and
interaction parameters, as well as chemical potentials. For example, under
periodic driving of this model coherent destruction of tunneling and dark
Floquet state have been predicted in theory \cite{Luo}. Thus, dark states
can also be controlled and realized in the three-state (three-well) system.
Even chaotic phenomena and bifurcation mechanism causing self-trapping have
been studied in the dynamics of three coupled condensate systems \cite%
{Penna,Penna2}. In addition, in the light propagation in waveguides, the
Kerr nonlinear interactions induce a variety of interesting quantum
phenomena \cite{Scully}. Thus when dissipation and the nonlinear interaction
together play the roles in triple-well system, novel features may be
expected in the dynamical evolution of the system.

In the present paper, we mainly study the quantum dynamics of a
non-Hermitian triple-well system. We focus on the time evolutions of modulus
squared of coefficients in three local states without nonlinear
interactions. The analytic solutions of the Schr\"{o}dinger equations
directly give the time-dependent information for uniform chemical potentials
in two end wells. The finding is that the eigenstate is a dark state, whose
eigenergy is zero, the projection on which is not dependent of time and the
loss factor. But when chemical potentials in two ends are not uniform, the
dark state would be not any more the eigenstate. Moreover, when nonlinear
terms in three wells are considered, the modulus squared in three wells will
be quickly diminished. These results are still suitable for systems of odd
wells with similar structure.

\section{Model and analytic solutions in linear case}

We consider a coupled triple-well system with an imaginary chemical
potential in the middle well. In general, the wave function $\left\vert \psi
\left( t\right) \right\rangle $ of the system is a superposition of states
at three local sites, i.e.,
\begin{equation}
\left\vert \psi \left( t\right) \right\rangle =c_{1}\left( t\right)
\left\vert 1\right\rangle +c_{2}\left( t\right) \left\vert 2\right\rangle
+c_{3}\left( t\right) \left\vert 3\right\rangle ,  \label{ansatz}
\end{equation}%
where $c_{i}(t)$ are the amplitudes for three states $\left\vert
i\right\rangle $ ($i=1,2,3$). In this local site space, where the spatial
dependence of the states will not be considered, the dynamic equation of the
system \cite{Livi,Ng,Konotop1} reads ($\hbar =1$)
\begin{equation}
i\frac{\partial }{\partial t}\left(
\begin{array}{c}
c_{1} \\
c_{2} \\
c_{3}%
\end{array}%
\right) =H\left(
\begin{array}{c}
c_{1} \\
c_{2} \\
c_{3}%
\end{array}%
\right)  \label{sch}
\end{equation}%
with the Hamiltonian%
\begin{equation}
H=\left(
\begin{array}{ccc}
\mu _{1}+g_{1}\left\vert c_{1}\right\vert ^{2} & -J & 0 \\
-J & \mu _{2}+g_{2}\left\vert c_{2}\right\vert ^{2} & -J \\
0 & -J & \mu _{3}+g_{3}\left\vert c_{3}\right\vert ^{2}%
\end{array}%
\right) .  \label{H}
\end{equation}%
The chemical potentials $\mu _{1}$ and $\mu _{3}$ are real and $\mu
_{2}=\eta -i\alpha $ is a complex number, which denotes an effective loss ($%
\alpha >0$) or a gain\ ($\alpha <0$) at the state $\left\vert 2\right\rangle
$ \cite{Wunner,Niederle}. $g_{i}$ is the strength of the Kerr nonlinearity
in state $\left\vert i\right\rangle $ and $J$ is the coupling strength \cite%
{Kottos2}. We set $J=1$ so that all energies are in units of $J$.

We first focus on the simplest case that the chemical potentials are
symmetrically distributed ($\mu _{1}=\mu _{3}$), and the interactions are
neglected $g_{i}=0$. The Schr\"{o}dinger equation (\ref{sch}) can be solved
by a substitution $c_{i}\left( t\right) =c_{i}^{0}\exp \left( -i\lambda
_{i}t\right) $ and one has the eigenvalues $\lambda _{i}$ for Hamiltonian (%
\ref{H})
\begin{align}
\lambda _{1}& =\mu _{1},  \notag \\
\lambda _{2,3}& =\frac{1}{2}\left( \mu _{1}+\mu _{2}\pm \Theta \right) ,
\end{align}%
and the corresponding ket space is spanned by three eigenvectors%
\begin{equation}
\left\vert \psi _{1}\right\rangle =\left(
\begin{array}{c}
-1 \\
0 \\
1%
\end{array}%
\right) ,\left\vert \psi _{2}\right\rangle =\left(
\begin{array}{c}
1 \\
\lambda _{3}-\mu _{2} \\
1%
\end{array}%
\right) ,\left\vert \psi _{3}\right\rangle =\left(
\begin{array}{c}
1 \\
\lambda _{2}-\mu _{2} \\
1%
\end{array}%
\right) ,  \label{Vector}
\end{equation}%
where $\Theta =\sqrt{\left( \mu _{1}-\mu _{2}\right) ^{2}+8}$. We do not
bother to normalize them because the normalization factor will not affect
the final result. The dual, bra space with eigenvector, e.g., $\left\langle
\psi _{2}\right\vert =(1,\lambda _{3}^{\ast }-\mu _{2}^{\ast },1)$ is not
orthogonal to the ket space. It is thus necessary to define the Hilbert
space of $H^{\dagger }$, $\left\vert \tilde{\psi}_{i}\right\rangle
=\left\vert \psi _{i}^{\ast }\right\rangle $, the bra vectors being $%
\left\langle \tilde{\psi}_{i}\right\vert =\left\langle \psi _{i}^{\ast
}\right\vert $. Here the symbol $\ast $ means the complex conjugate for all
complex numbers. These eigenvectors together form a biorthogonal basis, i.e.
the completeness relation reads \cite{Heiss}
\begin{equation}
\sum_{k}\frac{\left\vert \psi _{k}\right\rangle \left\langle \tilde{\psi}%
_{k}\right\vert }{\left\langle \tilde{\psi}_{k}|\psi _{k}\right\rangle }=1,
\end{equation}%
and the orthogonality means
\begin{equation}
\frac{\left\langle \psi _{k}|\tilde{\psi}_{k^{\prime }}\right\rangle }{%
\left\langle \psi _{k}|\tilde{\psi}_{k}\right\rangle }=\delta _{kk^{\prime
}}.
\end{equation}%
Note that the eigenvector $\psi _{1}$ is a dark state which is the
superposition of two local states in the left and right wells. The
completeness dictates that an arbitrary normalized initial state $\left\vert
\psi \left( 0\right) \right\rangle =\left(
c_{1}^{0},c_{2}^{0},c_{3}^{0}\right) ^{T}$ can be expressed in the
eigenvector space (\ref{Vector})
\begin{equation}
\left\vert \psi \left( 0\right) \right\rangle =A_{1}\left\vert \psi
_{1}\right\rangle +A_{2}\left\vert \psi _{2}\right\rangle +A_{3}\left\vert
\psi _{3}\right\rangle ,
\end{equation}%
where the coefficients $A_{i}$ are suitable combinations of $c_{i}^{0}$, for
example, $A_{1}=\left( c_{3}^{0}-c_{1}^{0}\right) /\sqrt{2}$ for a
normalized $\left\vert \psi _{1}\right\rangle $. At time $t$ the wave
function evolves according to
\begin{equation}
\left\vert \psi \left( t\right) \right\rangle =A_{1}e^{-i\lambda
_{1}t}\left\vert \psi _{1}\right\rangle +A_{2}e^{-i\lambda _{2}t}\left\vert
\psi _{2}\right\rangle +A_{3}e^{-i\lambda _{3}t}\left\vert \psi
_{3}\right\rangle .
\end{equation}%
The matrix for the time evolution operator $C$ in the local site space
defined as \cite{Konotop}
\begin{equation}
\left\vert \psi \left( t\right) \right\rangle =\left(
\begin{array}{ccc}
C_{11}\left( t\right) & C_{12}\left( t\right) & C_{13}\left( t\right) \\
C_{21}\left( t\right) & C_{22}\left( t\right) & C_{23}\left( t\right) \\
C_{31}\left( t\right) & C_{32}\left( t\right) & C_{33}\left( t\right)%
\end{array}%
\right) \left\vert \psi \left( 0\right) \right\rangle ,  \label{tt}
\end{equation}%
can be calculated as
\begin{equation}
C=S^{\dagger }Diag\left( e^{-i\lambda _{1}t},e^{-i\lambda
_{2}t},e^{-i\lambda _{3}t}\right) \tilde{S}
\end{equation}%
where $S$ is transformation matrix between the site space and the
eigenvector space with the adjoint matrix $\tilde{S}$. These operators are
necessarily not unitary. Actually, formed by merging the eigenvectors $|\psi
_{i}\rangle $ or $\left\vert \tilde{\psi}_{i}\right\rangle $ into a square
matrix row by row, the operators $S^{\dagger }$ and ${\tilde{S}}^{\dagger }$
satisfy $S^{\dagger }{\tilde{S}}={\tilde{S}}S^{\dagger }=1$ and ${\tilde{S}}%
^{\dagger }S=S{\tilde{S}}^{\dagger }=1$. In this way, the time-dependent
matrix elements $C_{ij}\left( t\right) $ in eq. (\ref{tt}) can be determined
as
\begin{align}
C_{11}\left( t\right) & =C_{33}\left( t\right) =\frac{1}{2}e^{-i\lambda
_{1}t}+\frac{1}{2}f_{+}\left( t\right) ,  \notag \\
C_{31}\left( t\right) & =C_{13}\left( t\right) =-\frac{1}{2}e^{-i\lambda
_{1}t}+\frac{1}{2}f_{+}\left( t\right) ,  \notag \\
C_{22}\left( t\right) & =f_{-}\left( t\right) ,  \notag \\
C_{12}\left( t\right) & =C_{21}\left( t\right) =C_{23}\left( t\right)
=C_{32}\left( t\right)  \notag \\
& =\frac{1}{\Theta }\left( e^{-i\lambda _{3}t}-e^{-i\lambda _{2}t}\right) ,
\end{align}%
where%
\begin{align}
f_{+}\left( t\right) & =\frac{\lambda _{2}-\mu _{2}}{\Theta }e^{-i\lambda
_{2}t}-\frac{\lambda _{3}-\mu _{2}}{\Theta }e^{-i\lambda _{3}t},  \notag \\
f_{-}\left( t\right) & =\frac{\lambda _{2}-\mu _{2}}{\Theta }e^{-i\lambda
_{3}t}-\frac{\lambda _{3}-\mu _{2}}{\Theta }e^{-i\lambda _{2}t}.
\end{align}%
The symmetry in the coefficients $C_{ij}$ reflects directly that of the
Hamiltonian in the local site space. For an arbitrary initial state $%
\left\vert \psi \left( 0\right) \right\rangle $, we can infer the state
evolution $\left\vert \psi \left( t\right) \right\rangle $ by the
coefficients $C_{ij}\left( t\right) $, i.e. the amplitude on the $i$-th
local state reads $c_{i}\left( t\right) =C_{i1}\left( t\right)
c_{1}^{0}+C_{i2}\left( t\right) c_{2}^{0}+C_{i3}\left( t\right) c_{3}^{0}$.
Correspondingly the modulus squared of coefficient in each local state are%
\begin{equation}
P_{i}\left( t\right) =\left\vert c_{i}\left( t\right) \right\vert ^{2},
\label{Pro}
\end{equation}%
and the sum is%
\begin{equation}
P_{all}\left( t\right) =\sum_{i}P_{i}\left( t\right) .
\end{equation}

\begin{figure}[tbp]
\includegraphics[width=0.45\textwidth]{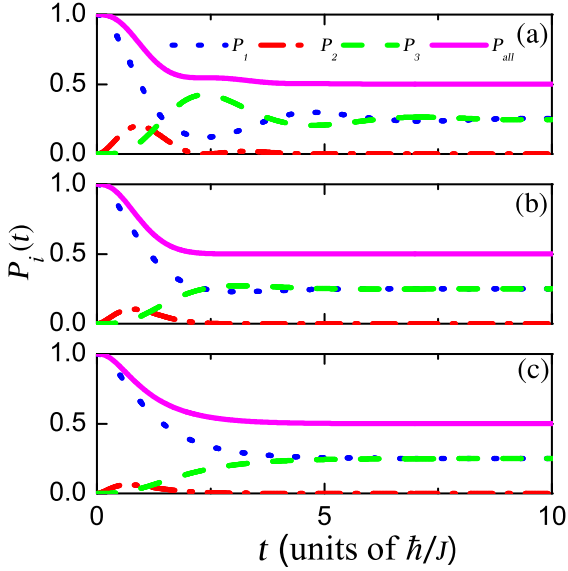}
\caption{Time evolution of the modulus squared of coefficient in each well $%
P_{i}(t)$ and the sum $P_{all}$ (pink solid line) with the loss factor $%
\protect\alpha =1$ (a), $\protect\alpha =2$ (b) and $\protect\alpha =3$ (c)
starting from the initial state $c_{1}^{0}=1$, $c_{2}^{0}=c_{3}^{0}=0$. Here
$\protect\mu _{1}=\protect\mu _{3}=\protect\eta =0$.}
\end{figure}

For simplicity now we assume that the chemical potentials in left and right
states vanish and that in local state $\left\vert 2\right\rangle $ is pure
imaginary, i.e. $\mu _{1}=\mu _{3}=0$ and $\mu _{2}=-i\alpha $. We focus on
the case of dissipation in the rest of the paper, i.e., $\alpha >0$. It is
easy to see that while $\Theta $ in the eigenvalues $\lambda _{2,3}$ is
positive and real for $\alpha ^{2}<8$, it becomes pure imaginary for $\alpha
^{2}>8$. This will greatly change the time dependence of the functions $%
f_{\pm }$. When $\alpha ^{2}<8$, we see $\Theta =\left\vert \Theta
\right\vert $ and the functions $f_{\pm }\left( t\right) $ reduces to%
\begin{equation}
f_{\pm }\left( t\right) =\pm e^{-\frac{1}{2}\alpha t}\frac{\sin \left( \frac{%
1}{2}\left\vert \Theta \right\vert t\pm \beta \right) }{\sin \beta },
\label{f1}
\end{equation}%
where $\beta =\arcsin \left( \left\vert \Theta \right\vert /\sqrt{8}\right) $%
. Due to the sinusoidal functions in Eq. (\ref{f1}), it describes an
oscillating decay process starting from $f_{\pm }(0)=1$. We find the
critical damping occurs at $\alpha ^{2}=8$. In the other case when $\alpha
^{2}>8$, $\Theta =i\left\vert \Theta \right\vert $, the system enters the
over-damping region
\begin{equation}
f_{\pm }\left( t\right) =\pm e^{-\frac{1}{2}\alpha t}\frac{\sinh \left(
\frac{1}{2}\left\vert \Theta \right\vert t\pm \beta ^{\prime }\right) }{%
\sinh \beta ^{\prime }},  \label{f2}
\end{equation}%
where $\beta ^{\prime }=\text{arcsinh}\left( \left\vert \Theta \right\vert /%
\sqrt{8}\right) $. Since $\left\vert \alpha \right\vert >\left\vert \Theta
\right\vert $, the decaying term $e^{-\alpha t/2}$ of $f_{+}\left( t\right) $
in Eq. (\ref{f2}) will be compensated by the monotonically increasing
hyperbolic sine function, which leads to over-damping, i.e. a relatively
slow decay compared with Eq.(\ref{f1}) in the chemical parameter region $%
\alpha ^{2}<8$. As an example, starting from the initial state $c_{1}^{0}=1$
and $c_{2}^{0}=c_{3}^{0}=0$, the distributions in three states are
respectively
\begin{align}
P_{1}\left( t\right) & =\left\vert C_{11}\left( t\right) \right\vert ^{2}=%
\frac{1}{4}\left\vert f_{+}\left( t\right) +1\right\vert ^{2},  \notag
\label{P} \\
P_{3}\left( t\right) & =\left\vert C_{31}\left( t\right) \right\vert ^{2}=%
\frac{1}{4}\left\vert f_{+}\left( t\right) -1\right\vert ^{2},
\end{align}%
and%
\begin{align}
P_{2}\left( t\right) & =\left\vert C_{21}\left( t\right) \right\vert ^{2}
\notag \\
& =\frac{2}{\left\vert \Theta \right\vert ^{2}}e^{-\alpha t}\left\{
\begin{array}{c}
1-\cos \left( \left\vert \Theta \right\vert t\right) ,\qquad \alpha ^{2}<8,
\\
\cosh \left( \left\vert \Theta \right\vert t\right) -1,\qquad \alpha ^{2}>8.%
\end{array}%
\right.
\end{align}%
\begin{figure}[tbp]
\includegraphics[width=0.45\textwidth]{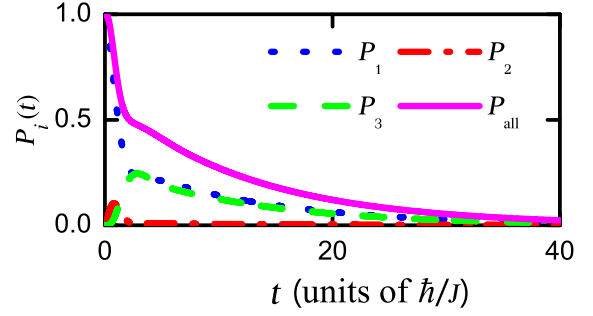}
\caption{Time evolution of the modulus squared of coefficient in each well $%
P_{i}(t)$ and the sum $P_{all}$ (pink solid line) for non-equal distribution
of the chemical potentials starting from the initial state $c_{1}^{0}=1$, $%
c_{2}^{0}=c_{3}^{0}=0$. Here $\protect\mu _{1}=0.1$, $\protect\mu _{3}=0.5$
and $\protect\alpha =2$. }
\end{figure}

The time evolution of the modes are shown in Fig. 1 for three different
values of $\alpha $. For $\alpha =1$, we find $P_{1}\left( t\right) $ and $%
P_{3}\left( t\right) $ undergo explicit oscillations around the same
equilibrium value $0.25$. These modulus squared values return rapidly to
their equilibrium value for $\alpha =2$, with an obvious damping
oscillation. For $\alpha =3$, however, the decrease of $P_{1}\left( t\right)
$ and the increase of $P_{3}\left( t\right) $ are much slower, which shows
the typical behavior of over-damping. In both cases $P_{2}\left( t\right) $
quickly oscillates to a vanishingly small value, which means the leakage of
the mode from the middle well. In addition we find the equilibrium values
for $P_{i}$ in the limit $t\rightarrow +\infty $ are independent of $\alpha $%
. In this limit $e^{-i\lambda _{2,3}t}\rightarrow 0$ and $f_{\pm }\left(
t\right) \rightarrow 0$, for arbitrary initial state the wave function $%
\left\vert \psi \left( t\right) \right\rangle $ reduces to%
\begin{equation}
\left\vert \psi \left( t\rightarrow +\infty \right) \right\rangle =\frac{%
c_{3}^{0}-c_{1}^{0}}{2}\left(
\begin{array}{c}
-1 \\
0 \\
1%
\end{array}%
\right) ,  \label{sta}
\end{equation}%
and the associated norms are%
\begin{align}
P_{1,3}\left( t\rightarrow +\infty \right) & =\frac{\left(
c_{1}^{0}-c_{3}^{0}\right) ^{2}}{4},\qquad  \notag \\
P_{2}\left( t\rightarrow +\infty \right) & =0.  \label{coh}
\end{align}%
This indicates that the steady state is the dark state $\left\vert \psi
_{1}\right\rangle $, the projection on which would be stored forever. For
the initial state $c_{1}^{0}=1$ and $c_{2}^{0}=c_{3}^{0}=0$, it is easy to
show that the total norm $P_{all}\left( t\rightarrow +\infty \right) =0.5$,
which does not vary with $\alpha $ as depicted in Fig. 1. These results show
apparent suppression of dissipation because the projection on the dark state
does not change with time. Similar things happen for other initial states,
even for the dark state $\psi _{1}$ itself \cite{Bender2,Baker,Graefe,Hao}.

The linear non-Hermitian system with non-zero chemical potentials can be
solved readily by $\left\vert \psi \left( t\right) \right\rangle =\exp
\left( -iHt\right) \left\vert \psi \left( 0\right) \right\rangle $ and the
three modulus squared parameters $P_{i}$ are given by (\ref{Pro}). For
non-equal chemical potentials in the left and right wells $\mu _{1}\neq \mu
_{3}$, the dark state $\left\vert \psi _{1}\right\rangle $ is not any more
the eigenstate of the system \cite{Dark}. An immediate result is that $P_{i}$
in all three states will be lost in the limit $t\rightarrow +\infty $. We
show this full leakage in Fig. 2 for $\alpha =2$, $\mu _{1}=0.1$ and $\mu
_{3}=0.5$. Clearly, the probability $P_{1}=1$ in the initial state $%
(1,0,0)^{T}$ decays from $1$ to $0$, at the same time $P_{2,3}$ reduce to
zero after a temporary increase.

Under the balanced condition $\mu _{1}=\mu _{3}$, it is interesting to study
the influence of the real part $\eta $ of $\mu _{2}$ on the evolution of
each state. As an example, we set $\mu _{1}=\mu _{3}=0.5$ and $\alpha =2$
and increase the real part $\eta $ from $0$ to $6$. The modulus squared $%
P_{1,3}$ are found to oscillate in longer and longer time, which effectively
slows down the process for the sum of $P_{i}$ to reach the equilibrium.
However, it has no effect on the distribution of the steady state in the
limit $t\rightarrow +\infty $.

\begin{figure}[tbp]
\includegraphics[width=0.45\textwidth]{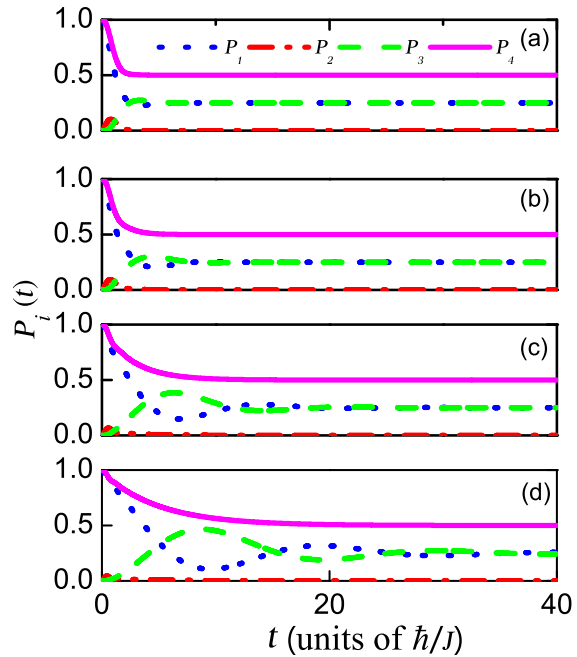}
\caption{The effect of the real part of $\protect\mu _{2}$ on the evolution
of $P_{i}\left( t\right) $ with $\protect\eta =0$ (a), $2$ (b), $4$ (c) and $%
6$ (d). Here $\protect\mu _{1}=\protect\mu _{3}=0.5$ and $\protect\alpha =2$%
. }
\end{figure}

\section{Numerical Scheme for nonlinear interaction case}

The analytical solution in above section is not available when the nonlinear
interaction is introduced in the Hamiltonian (\ref{H}). In looking for the
similar variational ansatz solution (\ref{ansatz}), we need to take several
approximations into account: (a) First of all, the time evolution of the
wave function (\ref{ansatz}) is described as the superposition of three
local states \cite{Smerzi}. The nonlinear terms in the dynamic equations (%
\ref{sch}), on the other hand, would destroy such a superposition. When the
probability in the tunneling region of the adjacent wells is small enough
such that the nonlinear interaction in these regions is negligible, the
superposition ansatz (\ref{ansatz}) is applicable. (b) In the meantime we
decompose the time and the spatial dependence of the wave function $%
\left\vert \psi \left( t\right) \right\rangle $, which has been verified
numerically in the study of BEC trapped in double well potential \cite{PHD}.
(c) The spatial dependence of the local states will not be considered here,
despite that the overlap of the states determines the tunneling strength $J$
and the interaction parameters $g_{i}$ \cite{Malomed1}. To investigate the
dynamics of the system with nonlinear terms, we deal with the time-dependent
Hamiltonian in the site space by means of the successive iteration, i.e.
starting from an arbitrarily normalized initial state $\left\vert \psi
\left( 0\right) \right\rangle $, the wave function at time $t+\delta t$ is
evolved from previous time $t$ through
\begin{equation}
\left\vert \psi \left( t+\delta t\right) \right\rangle =\exp \left(
-iH[t]\delta t\right) \left\vert \psi \left( t\right) \right\rangle ,
\label{EV}
\end{equation}%
while the time-dependence of the Hamiltonian $H[t]$ is described by the
interaction terms $\left\vert c_{i}\left( t\right) \right\vert ^{2}$ in
equation (\ref{H}). Accordingly, we numerically split the evolution time $t$
into many small intervals with the time step $\delta t$ being small enough
to admit a solution with good precision. We note that unlike in the case of
time-dependent Gross-Pitaevskii equations for dissipative BEC in double well
\cite{Hao} or the barrier transmission of BEC in waveguide \cite{Graefe},
the absence of the kinetic term makes it much easier for the convergence of
the solution.

We now discuss the typical numerical results with different nonlinear
parameters. For nonzero interaction existing only in the middle well $%
g_{1}=g_{3}=0$ and $g_{2}=3$, the stationary solutions for different loss $%
\alpha =1$ and $3$ are shown in Fig. (4a) and (4b). The stationary solutions
are identical to results of noninteracting case in Fig. (1b) and (1c), which
shows that $g_{2}$ does not affect the evolution of $P_{i}\left( t\right) $
in limit $t\rightarrow \infty $. For three identical interaction parameters $%
g_{1}=g_{3}=g_{2}=3$, on the other hand, we observe quite different
behavior. The nonlinear terms $g_{1,3}$ obviously diminish the projection of
the dark state to a very low level, and moreover $P_{i}$ also decrease with
the increase of $\alpha $. Unequal $g_{1}\neq g_{3}$ would destroy the
coherent character completely, leading to a full leakage of the wave-packet.
\begin{figure}[tbp]
\includegraphics[width=0.45\textwidth]{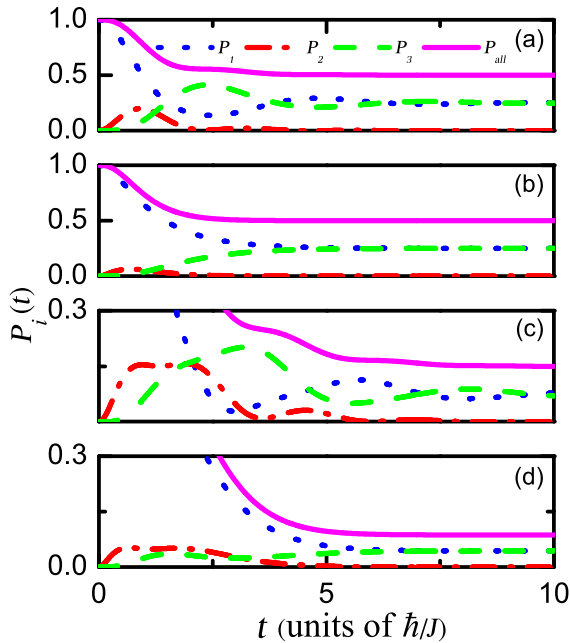}
\caption{A numerical investigation of evolutions for different nonlinear
parameters: When $g_{1}=g_{3}=0$, $g_{2}=3$, $\protect\alpha =1$ (a) and $3$
(b); When $g_{1}=g_{3}=g_{2}=3$, $\protect\alpha =1$ (c) and $3$ (d). Here $%
\protect\mu _{1}=\protect\mu _{3}=0$ and initial conditions: $c_{1}^{0}=1$
and $c_{2}^{0}=c_{3}^{0}=0$.}
\label{fig4}
\end{figure}

\section{Generalization to any odd site number}

We have dealt with the three-well model with a loss in the middle site and
found there is a dark state when $\mu _{1}=\mu _{3}$ and $g_{i}=0$. Then
based on this we also discussed the time evolution for different parameters
and even different interactions. These results can be generalized to a
general $(2n+1)$-well system where only the middle site has a loss and is
coupled with other wells. For simplicity, firstly we consider the five-well
model. The Hamiltonian is%
\begin{equation}
H_{5}=\left(
\begin{array}{ccccc}
\mu _{1} & 0 & -1 & 0 & 0 \\
0 & \mu _{2} & -1 & 0 & 0 \\
-1 & -1 & \mu _{3} & -1 & -1 \\
0 & 0 & -1 & \mu _{4} & 0 \\
0 & 0 & -1 & 0 & \mu _{5}%
\end{array}%
\right)  \label{H5}
\end{equation}%
where $\mu _{3}=-i\alpha $ and $\mu _{i\neq 3}=0$ and only $\left\vert
3\right\rangle $ is coupled with the rest of wells. When dealing with $%
\left\vert H_{5}-EI\right\vert =0$, we can obtain%
\begin{equation*}
E^{3}\left( E^{2}+i\alpha E-4\right) =0.
\end{equation*}%
For eigen energy $E=0$, we can get%
\begin{eqnarray}
c_{3} &=&0,  \notag \\
c_{1}+c_{2}+c_{4}+c_{5} &=&0,  \label{five}
\end{eqnarray}%
where $c_{i}$ are the coefficients of $\left\vert i\right\rangle $. The
solutions of (\ref{five}) are not unique and correspond to triplet dark
states with a node structure in the middle well and other coefficients
summed up to zero. Hence we analyze the dynamics of this model by the
numerical method, just as the section. III in this paper. But from (\ref%
{five}) we can find that the eigenvectors with $E=0$ have not projection in $%
\left\vert 3\right\rangle $ because of $c_{3}=0$, which is independent of
any parameter. So the projection in these eigenvectors would not vary over
time. In order to further explain it, using numerical method (\ref{EV}) we
can find when $\left\vert \psi \left( 0\right) \right\rangle =\left(
1,0,0,0,0\right) ^{T}$,%
\begin{equation*}
\lim_{t\rightarrow \infty }P_{all}\left( t\right) =3/4,
\end{equation*}%
which is the projection in the eigenvectors with $E=0$ and is much larger
than in three-well system as shown in Fig. 5. Accordingly, we study an
arbitrary $(2n+1)$-well system, and set $\mu _{n+1}=-i\alpha $ and $\mu
_{j\neq n+1}=0$. With the same numerical method we find the law: $%
\lim_{t\rightarrow \infty }P_{all}\left( t\right) =1-1/2n$ with $\left\vert
\psi \left( 0\right) \right\rangle =\left( 1,0,0,\cdots ,0\right) ^{T}$,
which coincides the above-mentioned results. This gives a good application
that more wells may be used to construct sophisticated dark states, the
projection onto which is kept on a much higher level in the steady state of
dynamical evolution.
\begin{figure}[tbp]
\includegraphics[width=0.45\textwidth]{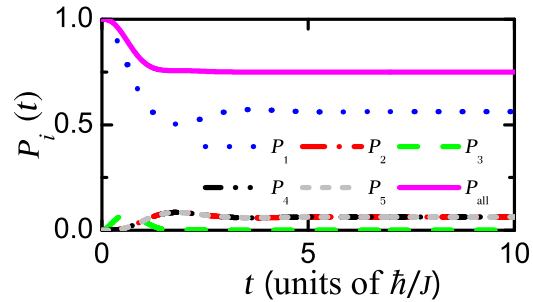}
\caption{Time evolutions of modulus squared of coefficients in five wells
for initial states $\left\vert 1,0,0,0,0\right\rangle $. Here nonlinear
terms $g_{i}=0$ and the loss factor $\protect\alpha =2$.}
\label{fig5}
\end{figure}

\section{Conclusions}

We have presented a detailed analysis of dynamical evolutions of the
three-well system with a loss in the middle well. When the chemical
potentials in two end wells are real and uniform, there is always a dark
state,\textit{\ }the projection on which does not change over time and the
loss factor $\alpha $. But there exists a critical value, where the norms at
two end sites evolute from damping oscillation to over-damping. When the
chemical potentials in two end wells are not uniform, the dark state is not
any more the eigenstate of the system and three norms will decay to zero. In
addition, when the nonlinear interactions are introduced and uniform in two
end wells, the projection of dark state will be obviously diminished, but do
not disappear. And the projection also decrease with the increase of the
loss factor. However, the interaction at middle site plays no role and the
dark state is proven to be the key to the suppression of the dissipation. In
addition, the other two interaction intensities would promote the loss. When
extending the triple-well system to a general $(2n+1)$-well we found that
the total norm follows the law $1-1/2n$ in the absent of interactions, which
can be used to enhance the anti-leakage capability in signal propagation in
certain medium with dissipation.

This work is supported by the NSF of China under Grants No.11574187,
11674201, 11474189, 11425419 and 11374354.

\end{document}